\documentclass[a4paper, amsfonts, amssymb, amsmath, reprint, showkeys, nofootinbib, twoside,superscriptaddress,aps,prapplied,floatfix]{revtex4-2}

\usepackage[utf8]{inputenc}
\usepackage[colorinlistoftodos, color=green!40, prependcaption]{todonotes}

\begin{document}
\title{Modified Split Ring Resonators for Efficient and Homogeneous Microwave Control of Large Volume Spin Ensembles}

\author{Yachel Ben-Shalom}
\email{yachel.b.s@gmail.com}
\affiliation{Dept. of Applied Physics, The Hebrew University of Jerusalem, Jerusalem, Israel.}
\affiliation{ELTA systems Ltd. Ashdod, Israel.}

\author{Amir Hen}
\affiliation{Dept. of Applied Physics, The Hebrew University of Jerusalem, Jerusalem, Israel.}

\author{Nir Bar-Gill}
\affiliation{Dept. of Applied Physics, The Hebrew University of Jerusalem, Jerusalem, Israel.}
\affiliation{The Racah Institute of Physics, The Hebrew University of Jerusalem, Jerusalem, Israel.}
\date{\today}

\begin{abstract}
Quantum sensing using local defects in solid-state systems has gained significant attention over the past several years, with impressive results demonstrated both in Academia and in Industry. Specifically, employing large volume and high density ensembles for beyond state-of-the-art sensitives is of clear interest. A major obstacle for achieving such record sensitivities is associated with the need to realize strong, homogeneous driving of the sensor defects. Here we focus on high-frequency microwave sensing using nitrogen-vacancy centers in diamond, and develop a modified split-ring resonator design to address this issue. We demonstrate enhanced drive strengths and homogeneities over large volumes compared to previous results, with prospects for enabling the desired sensitivites. We reach Rabi frequencies of up to 18 [MHz] with an efficiency ratio of 2 [$Gauss/\sqrt{Watt}$], along with an inhomogeneity of $<0.7\%$ in a volume of $0.1\:mm^3$.
\end{abstract}

\maketitle

\section{Introduction} \label{sec:Introduction}
In recent years, the interest in quantum metrology has been growing both in Academia and in Industry. A significant aspect of quantum metrology relates to magnetic sensing and specifically sensing of high frequency (GHz-range) alternating current (AC) magnetic fields, with several quantum systems being employed to enhance the sensitivity in this context \cite{degen2017quantum}. 

High frequency microwave sensing is a long-standing pillar of state-of-the-art applications, including radar, communication, and electron paramagnetic resonance. Specifically, sensing of weak, fast, transient signals (at the microsecond scale) is of extreme importance for both civilian and defense applications.

This work focuses on employing nitrogen-vacancy (NV) color center ensembles in diamond as the quantum sensor, specifically in the context of transient, high-frequency microwave signal detection. 

The NV center is a nanoscale defect in a diamond's crystal structure, and it can be initialized via optical pumping, read out via fluorescence intensity, and coherently manipulated using microwave (MW) fields. 

NV centers possess very high magnetic field sensitivity per unit volume \cite{taylor2008high, maze2008nanoscale, pham2012enhanced, acosta2010broadband}, yet realizing an ensemble sensor that can surpass state-of-the-art classical capabilities is still challenging. 
Leading classical sensors include semiconductor diodes \cite{radar2013electronic}, pickup coils (as are used in NMR and EPR) \cite{seton2005liquid}, and atomic magnetometers \cite{SAVUKOV2007214}.

While a full comparison of all of these sensors is beyond the scope of this work, we briefly note that for the high-frequency (GHz scale) transient (microsecond scale) signals of interest here, pickup coils and atomic magnetometers exhibit limited performance (while excelling at lower frequencies). Semiconductor diode sensors constitute the current state-of-the-art for radar applications, yet their noise floor characteristics are less promising than large spin ensemble sensors (a relevant comparison also with regard to ensemble Rydberg atom sensors \cite{santamaria2022comparison}).

Employing a large ensemble of NV centers naturally leads to a boost in sensitivity of the diamond sensor through a square root scaling in the number of defects \cite{wolf2015subpicotesla}, although exploiting this scaling to achieve practical record sensitivities requires overcoming several hurdles. One of them is achieving a MW driving field strong enough to cover the inhomogeneous broadening of the NV ensemble, while maintaining high homogeneity of the drive itself. Common approaches utilize pulsed dynamical decoupling (PDD) which is relatively robust to drive inhomogeneities \cite{genov2020efficient, farfurnik2015optimizing, ryan2010robust, naydenov2011dynamical}, yet it is based on low frequency ($<10\, MHz$) sensing protocols such that its applicability to high frequency sensing is limited. Protocols based on continuous dynamical decoupling (CDD) can achieve high-frequency ($0.1-10\, GHz$) sensing \cite{stark2017narrow, genov2019mixed, meinel2021heterodyne, wang2022picotesla, wang2022sensing, aharon2016fully}, but are much more sensitive to inhomogeneities. 

\begin{figure*}[tbh]
\begin{centering}
\includegraphics[width=0.95\textwidth]{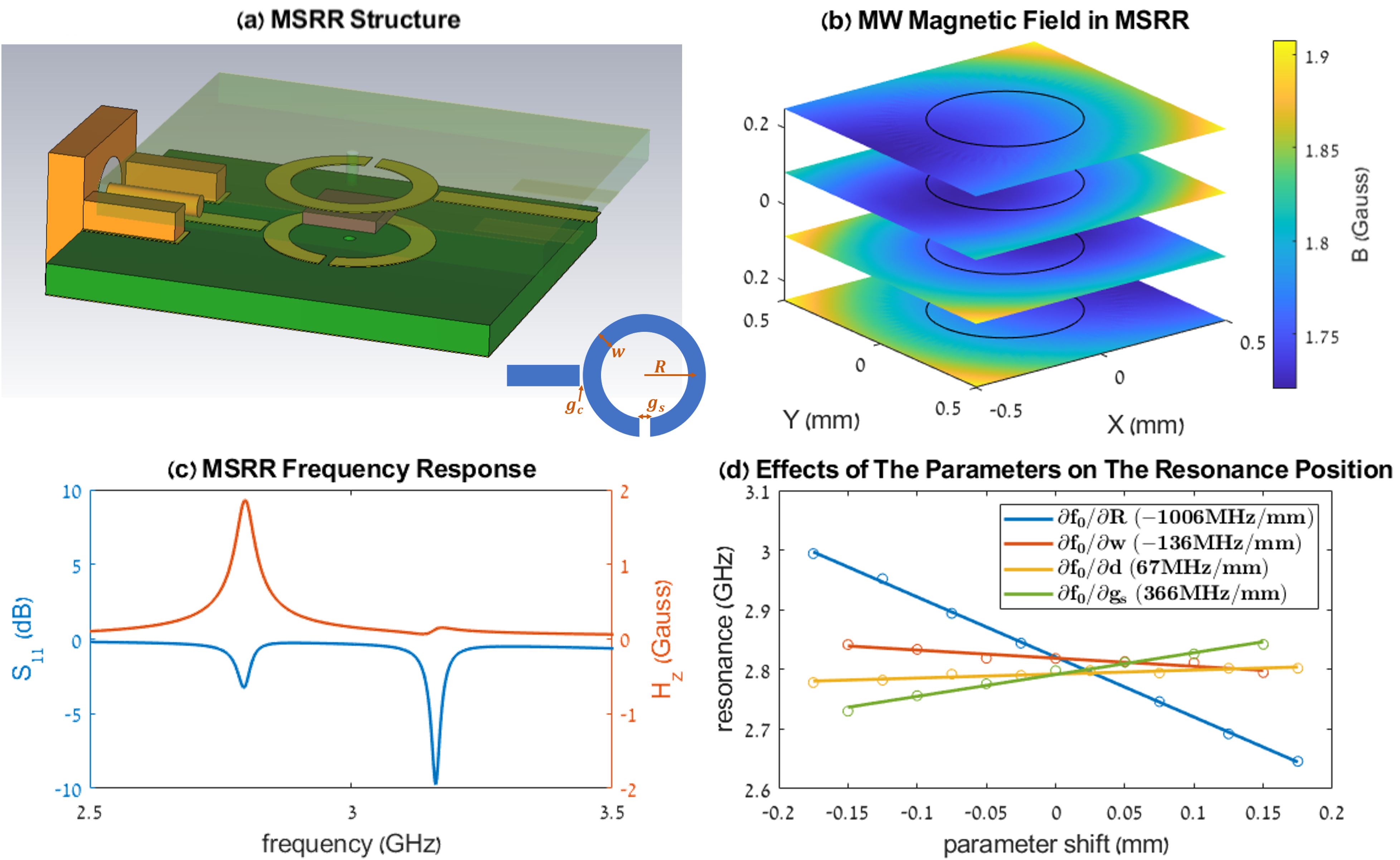}
\par\end{centering}
\caption{\label{fig: simulations}(a) CAD structure of MSRR. The chosen parameters are $R=2.9\:mm$, $w=1\:mm$, $g_s=0.4\:mm$, $g_c=0.1\:mm$, and the distance between the rings is $d=2.2\:mm$. (b) Simulated Z component of the MW magnetic field, which is the dominant and homogeneous in this structure. The results are for 1 Watt input and dual channel configuration. (c) Simulations of the $S_{11}$ vs. frequency, depicting two resonances - symmetric and anti-symmetric. Just in the symmetric resonance, the Z component of the magnetic field is high, when the currents in the loops are in the same directions. (d) Simulations of the effect of the different parameters on the resonance position.}
\end{figure*}

During the last few years, several solutions have been presented to overcome the challenge of MW control on large volumes. A common approach uses Double Split Ring Resonators (DSRR) \cite{bayat2014efficient, sturner2021integrated}, or different printed circuits \cite{wang2020integrated}, which have good homogeneity in the plane parallel to the resonator, but relatively low homogeneity along the normal axis. Other solutions with high 3D homogeneity suffer from low power \cite{alsid2022solid}, or require significant space around the diamond \cite{kapitanova20183d, wang2022picotesla, 20.500.12030_6372, takemura2022broadband, le2014addressing}. Such requirements pose major limitations on practical ensemble NV sensors, since they restrict efficient optical fluorescence collection needed for readout with high signal to noise ratio (SNR). 

In this paper we present a different microwave delivery design, based on structures used in the left-handed meta-materials field called Modified Split Ring Resonator (MSRR) \cite{marques2002role, marques2003comparative}. This structure consists of two split rings, but unlike the DSRR, the rings are placed one above the other. This approach provides high homogeneity also along the normal axis, with strong MW power density, and useful optical access through the spacing between the rings. This resonator can be designed to work at a wide range of frequencies, from $\sim 1\, GHz$ to $\sim 6\, GHz$.

Below we detail the design of the device and demonstrate its performance on a large volume ensemble of NVs.

\section{Simulations}

The design of the MSRR is depicted in Fig. \ref{fig: simulations}(a). We characterized the performance of the resonator numerically, and optimized the design parameters accordingly. The simulations were carried out using a commercial finite elements electromagnetic analysis package (CST). For each simulation we extracted the $S_{11}$ parameter (reflection coefficient \cite{balanis2015antenna}) and the magnetic field at the center along the normal axis ($\hat{z}$), as a function of the exciting frequency. 

We analyzed the strength and homogeneity of the microwave magnetic field by studying its behavior in a cylindrical volume, determined by the relevant laser excitation volume (assuming excitation through the rings) and the thickness of common diamond samples (up to $0.5$ mm). Therefore the effective volume is assumed to be a cylinder with a height of $0.5$ mm and a diameter varying between $0.1-1.5$ mm. For resonators with resonance frequencies around the NV resonance ($2.87\:GHz$), we extracted the full central magnetic field in this cylindrical volume [see Fig. \ref{fig: simulations}(b)]. We note that for the MSRR design we can either drive the MW field from a single port or from both ports simultaneously.

\begin{figure*}[!tbh]
\begin{centering}
\includegraphics[width=0.95\textwidth]{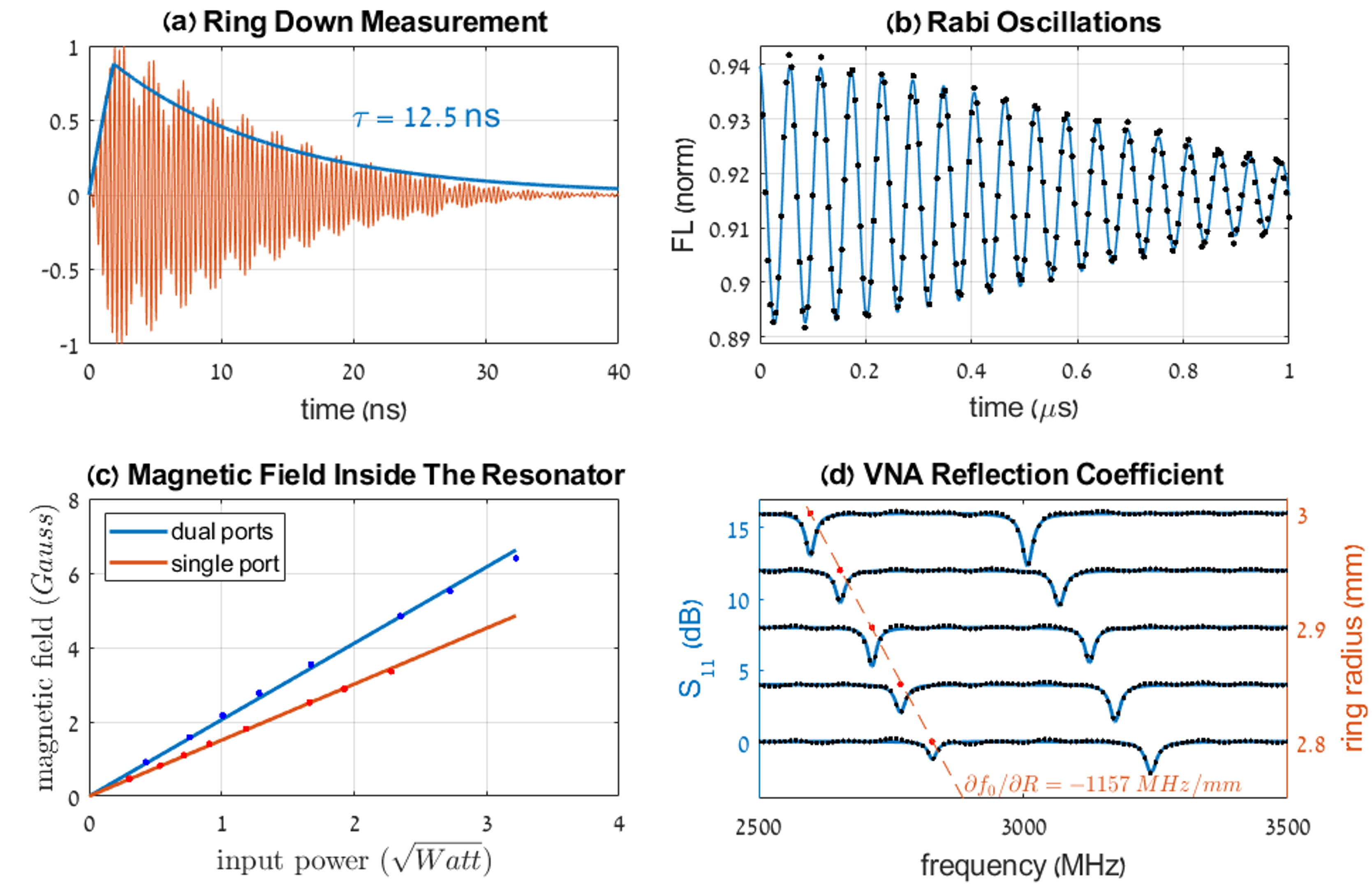}
\par\end{centering}
\caption{\label{fig: exp results} Experimental results: (a) Ring down measurement of the resonator (R=2.95 mm). The build-up time is $12.5 \pm 0.7\,ns$, a short enough time that enables efficient short pulses. (b) Rabi oscillations of NV center in MSRR. (c) The magnetic field inside the resonator as a function of the input power - dual port and single port (average of both separately). The ratio is 2(1.5) [Gauss/$\sqrt{Watt}$] in dual (single) port. (d) VNA reflection coefficient ($S_{11}$) for different resonators. The effect of the ring radius is $-1157\pm 12\: [MHz/mm]$, a bit greater than in the simulation.}
\end{figure*}

Our simulations show various parameter combinations leading to efficient MW excitation, and we choose optimal and practical design parameters detailed in Fig. \ref{fig: simulations}(a). The average magnetic field in the relevant effective cylindrical volume is $1.85 (1.31)$ [Gauss/$\sqrt{Watt}$] using dual (single) excitation ports. We define the inhomogeneity of the achieved microwave drive as the standard deviation of the magnetic field normalized by its mean value. We obtain an inhomogeneity of $0.43 (0.67) \%$ for a diameter of $0.5$ mm in dual (single) port driving [Fig \ref{fig: inhomogeneity}].  

We note that although the best results are obtained with simultaneous excitation of both ports (as expected), we found that even for single port excitation the achieved homogeneity is quite good, due to the strong coupling of the rings. This can be important because the calibration of power and phase for two channels driving can be challenging and impose additional noise and instabilities in practical scenarios.

In this type of resonator we naturally obtain two different resonances. The lower resonance is symmetric with both loops driven by current along the same direction, and the higher resonance is anti-symmetric with current driving in opposite directions in the two rings. We are interested in the symmetric resonance because the relevant drive component is the magnetic field along the normal axis. In Fig. \ref{fig: simulations}(c) we plot the $S_{11}$ simulations (blue) clearly depicting the two resonances, while the simulated magnetic field (red) results in one pronounced peak at the symmetric resonance.

We examined the effect of several parameters on the position of the symmetric resonance [Fig. \ref{fig: simulations}(d)]. The most significant parameter is the ring radius, leading to a sensitivity of the resonance to the radius of $-1006$ [MHz/mm]. Additional relevant design parameters are the distance between the rings with a sensitivity of $67$ [MHz/mm], the width of the rings' strip with a sensitivity of $-136$ [MHz/mm], and the width of the rings' slit with a sensitivity of $366$ [MHz/mm].

Another advantage of this structure is its low sensitivity to external perturbations. This is due to the fact that the structure is encapsulated between two ground plates. The dielectric components inside the resonator can cause a shift in the resonance position of a few tens of MHz, but once fixed, it remains stable. Then, dielectric and metal components outside of the resonator plates have a significantly lower effect on the resonance position, of fewer than $\sim 2$ MHz.

\section{Experiments}

We chose the material of the PCB board to be Rogers 6010.2LM due to its high dielectric constant. We printed the optimized design obtained from the simulations (see above), with a range of ring radii around the optimal value in order to obtain flexibility in the resonance position. 

Our characterization measurements of the MSRR design were performed using a custom-built scanning confocal microscope on a relevant, high NV density diamond sample. Optical excitation was provided by a $532\:nm$ diode-pumped solid-state (DPSS) laser (Laser Quantum axiom 532) focused onto the diamond using a 10\texttimes, 0.25 NA objective. The excitation laser was pulsed by focusing it through an acoustic-optical modulator (G\&H R15260). NV fluorescence was collected through the same objective and separated from the excitation beam using a dichroic filter (Semrock Di02-R635-25X36). The light was additionally filtered (Thorlabs high-pass FELH0650 and notch NF533-17) and focused onto a single-photon counting module (Excelitas Technologies SPCM-780-13-FC).

The resonator was driven by an amplified (Mini-circuits ZHL-16W-43-S+) MW generator (Windfreak synthHD) and modulated by a switch (Mini-circuits ZASW-2-50DRA+). Microwave and optical pulses were controlled using a computer-based digital delay generator (Swabian Instruments Pulse-Streamer). Measurement protocols (pulse sequences, data acquisition, etc.) were controlled by custom software.

The static magnetic field was applied with a permanent magnet whose distance and position relative to the NV center were controlled by three translation stages (theta, phi, rho, Standa 8MT30-50, and 8MR151). The diamond position was controlled by 3 piezo-electric stages (x, y, z, SmarAct SLC2430). The diamond sample used in these experiments was model DNV-B1 (Element Six), with an NV density of $\sim300\:ppb$. 

We first characterize our MSRR devices using a MW network analyzer (Agilent N5230A), measuring the resonance and S parameters. We extracted the Q factor from the $S_{11}$ curves, and found it to vary from 121 to 146 in different resonators, and the bandwidth to vary from 19 to 23 [MHz]. Ring-down measurements show a build-up time of $12.5 \pm 0.7 ns$, a short enough time to enable standard pulses required for sensing protocols [Fig. \ref{fig: exp results}(a)].

We then measure the achievable Rabi driving as a function of the delivered MW excitation power. A characteristic Rabi oscillation curve at a power of 9 [Watt] is depicted in Fig. \ref{fig: exp results}(b), while the magnetic field as a function of drive power is shown in Fig. \ref{fig: exp results}(c). The ratio we measured is 2(1.5) [Gauss/$\sqrt{Watt}$] in dual (single) port - a bit higher than in the simulation.
The measured behavior, specifically Figs. \ref{fig: exp results}(c,d), show good agreement with the simulations as a function of the different design parameters, including the resonance position, the effect of the ring radius, and the magnetic field amplitude. 

\begin{figure}[tb!h]
\begin{centering}
\includegraphics[width=1\linewidth]{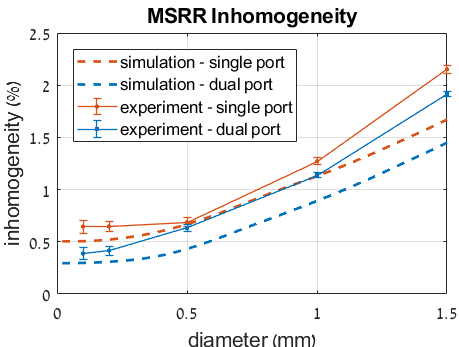}
\par\end{centering}
\caption{\label{fig: inhomogeneity} Inhomogeneity of the field in MSRR: simulation (dashed) and experiments (solid), dual port and single port excitation.}
\end{figure}

In order to analyze the field homogeneity we measure Rabi oscillations and extract the Rabi frequency at many points inside the resonator. For each radius, we took 8 points at equal distances, repeated for 4 points along the Z axis. Each point was measured both with dual port excitation and with a single port. Finally, we calculate the standard deviation for each radius and add a weight term that takes into account the different distances between the points. 
The measured inhomogeneity, Fig. \ref{fig: inhomogeneity}, agrees well with simulations and is quite low, with $<0.7\%$ inhomogeneity at a diameter of $0.5$ [mm] for both single and dual port excitations.

We note that while the results for single port excitation are close to the simulations (Fig. \ref{fig: inhomogeneity}), the dual port results are slightly less homogeneous compared to the simulations. We attribute this discrepancy to a misalignment of the loops or to a difference in the power delivered to the two channels, and we expect that it can be improved through more careful calibration. 

\section{Discussion}
A comparison of different designs for the same purpose \cite{takemura2022broadband} shows that the MSRR has excellent homogeneity and high power density, but it has a narrow bandwidth of around 20 MHz. While this bandwidth is suitable for a wide range of protocols that involve many frequencies around a certain resonance, it is not suitable for vector magnetometry based on the detection of multiple NV orientations.

Analyzing the effect of the MSRR on sensor performance is not trivial, as sensitivity is affected by many parameters. The optimal sensitivity of NV-based sensors is defined as \cite{stark2017narrow}:

\begin{equation}
\eta = \frac{1}{\gamma_{NV}} \frac{1}{\alpha C(T_2)} \frac{1}{\sqrt{N_{ph} T_2}} 
\label{eq:sensitivity}
\end{equation}

where $\gamma_{NV} = 28\, GHz/T$, $C(T_2)$ is the contrast after time $T_2$, $\alpha$ is a constant depending on the sensing protocol, $N_{ph}$ is the number of photons collected in one experimental run, and $T_2$ is the coherence time of the system.

Naively, when all the parameters are kept constant, the sensitivity of the sensor improves proportionally to the square root of the sensing volume. Additionally, when the limit on the coherence time comes from the MW spatial inhomogeneity, the sensitivity also grows proportionally to the inverse of the square root of the inhomogeneity. In real systems, the situation is more complicated. State-of-the-art systems for high-frequency sensing use a low drive field to obtain better homogeneity, and the sensitivity is typically around 5 pt/$\sqrt{\text{Hz}}$ \cite{alsid2022solid, wang2022picotesla}. However, a low-power drive reduces the contrast when it is below the inhomogeneous broadening, and it cannot be used for effective low-frequency noise decoupling. The MSRR enables an efficient drive, leading to enhanced sensor performance under certain assumptions.

We use Eq. \ref{eq:sensitivity} and estimate the sensitivity based on the different parameters. The contrast $C(0)$ of bulk NV diamond is usually around 5\%, and for simplicity, we can take $\alpha$ to be 1/4 for the double-drive protocol \cite{stark2017narrow}. 

The coherence time $\tau$ depends on the drive strength ($f_0$) and inhomogeneity ($\sigma$), but can be estimated in our case to be determined by the Fourier transform of the second(double) drive as $\tau=\left(\sqrt{2}\pi\sigma f_0 \right)^{-1} \approx 200 \mu$s.

The total number of photons collected per measurement can be estimated based on the excitation laser power, the NV density, excitation volume, saturation intensity and collection efficiency. For simplicity, we can assume an excitation laser power of $10\, Watt$, $10\%$ laser-fluorescence conversion ratio, and $0.3\,\mu$s detection window per measurement. These parameters are slightly lower than in Barry et al. \cite{barry2023sensitive}, and lead to an approximate value of $N_{ph} \approx 10^{12}$.

Following these estimates, the current approach could achieve a sensitivity of $\simeq 0.5 pT/\sqrt{Hz}$, which is an order of magnitude better than the current state-of-the-art results \cite{alsid2022solid, wang2022picotesla}.

 Regarding the collection path, using our resonator along with the standard collection method of a Compound Parabolic Collector (CPC) \cite{wolf2015subpicotesla} achieves decent results (as described above). However, in future work we will study an improved design of an optical collector with a narrow form factor, achieving estimated collection efficiencies above 50\%.

\section{Conclusions}
We have designed and demonstrated a modified split ring resonator (MSRR) device, aimed at achieving strong and homogeneous driving of large volume spin ensembles, while maintaining a form factor which allows for efficient optical coupling and fluorescence collection. This is motivated by a broad range of applications of quantum sensing, aiming to use optically active solid-state spin defects. Specifically, we focus on high-sensitivity sensing of high-frequency microwave fields using large ensembles of NV centers in diamond.

Our design significantly improves upon the current state of the art, achieving drive strengths of 17.2 [MHz] for a MW excitation power of 9 [Watt], with inhomogeneity below $0.7 \%$, for a cylindrical volume of diameter $0.5$ mm and height $0.5$ mm. Importantly, our design enables such results while maintaining close optical access for efficient fluorescence collection, which is a severe limitation in other approaches \cite{kapitanova20183d, takemura2022broadband}.

We expect that this design will enable beyond state-of-the-art sensitivities for high-frequency microwave signal sensing, push toward below $pT/\sqrt{Hz}$.

\begin{acknowledgments}
Y.B. acknowledges support from the Center for Nanoscience and Nanotechnology and ELTA Systems Ltd. 

N.B. acknowledges support from the European Union’s Horizon 2020 research and innovation program under grant agreements No. 101070546 (MUQUABIS) and No. 828946 (PATHOS), and has been supported in part by the Ministry of Science and Technology, Israel, the innovation authority (project \#70033), and the ISF (grants \# 1380/21 and 3597/21).

Also, the authors would like to thank Yuri Feldman, Nissan Maskil, Nimrod Teneh, and Ido Finkelman for fruitful discussions and suggestions.
\end{acknowledgments}

\bibliographystyle{apsrev4-2}
\bibliography{references.bib}

\end{document}